# Magneto-photoelectrochemical 2D heterojunction platform for biosensing detection

Tao Wang[1#], Nan Zhang[1#*], Hongjie Huang[4], Yunhe An[2], Yunyun Dai[5], Yongrui Li[1], Nan Yang[2], Chaojie Yang[2], Xinran Zhou[2], Yucheng Zhu[4], Yingshan Ma[5], Lingling Huang[1*], Yongtian Wang[1], Yang Liu[3*], Zhiyong Yan[2*]

**Abstract:** Photoelectrochemical (PEC) biosensors exhibit significant potential for biomolecule detection due to their high sensitivity and low background noise. However, their performance is severely constrained by the rapid recombination of photogenerated charge carriers. This study innovatively introduces a non-contact magnetic modulation strategy to suppress electron-hole recombination by manipulating carrier spin states, thereby significantly enhancing photoelectric conversion efficiency. Building on this mechanism, we developed a novel magnetically modulated PEC biosensing platform based on the MXenes/cobalt-doped titanium dioxide (Co-TiO$_2$) heterostructure. This platform achieved ultrasensitive detection of protein kinase A (PKA) activity. Compared to an identical probe-modified biosensor without magnetic field application, the developed platform demonstrated a 68.75% enhancement in detection sensitivity and achieved an ultralow detection limit for PKA of 0.00016 U/mL. It also exhibited a wide linear range from 0.005 to 80 U/mL. This research not only provides a novel methodology for kinase activity analysis but also pioneers the innovative strategy of magnetic modulation for enhanced PEC sensing. It opens new avenues for developing high-performance biosensing platforms, holding significant promise for early disease diagnosis and drug screening applications.

## Introduction

Photoelectrochemical (PEC) biosensors, leveraging significant advantages such as high sensitivity and low background noise, demonstrate considerable potential in biomolecule detection[1–3]. Their operational principle relies on the coupling of photoexcitation and electrochemical processes. When light with energy greater than or equal to the band gap of the semiconductor material irradiates the material, electrons in the valence band are excited and transition to the conduction band, simultaneously generating holes in the valence band, thereby forming photogenerated carriers (i.e., electron-hole pairs). These carriers migrate to the electrode or solution interface, where they initiate or modulate electrochemical reactions[4–6]. The real-time measurement of resulting electrical signals, such as photocurrent, photovoltage, or charge transfer quantity, enables highly sensitive and selective detection of target analytes[7]. However, the rapid recombination of photogenerated carriers constitutes a critical bottleneck constraining the enhancement of PEC sensor performance, severely limiting photoelectric conversion efficiency and detection sensitivity[8–10]. Current predominant strategies focus on suppressing recombination through intrinsic material structural modifications (e.g., elemental doping[11,12], noble metal deposition[13,14], and construction of heterojunction structures[15–18]). Although these approaches maximize the suppression of electron-hole pair recombination, sole reliance on intrinsic structural modifications faces inherent limitations for achieving further performance improvement[19].

To overcome the limitations of conventional material-centric approaches, the integration of structural modifications with external field modulation strategies (e.g., mechanical stress fields, thermal fields, electric fields, and magnetic fields) has emerged as an effective pathway for active PEC performance optimization[20,21]. These external fields offer significant advantages by altering intrinsic material properties or reaction environments. For instance, mechanical stress can modulate band structures and induce piezoelectric effects to promote charge separation[22,23]. Thermal fields accelerate reaction kinetics and improve carrier transport[24,25]. Electric fields serve as the primary driving force for carrier separation[26,27]. However, stress fields risk material fatigue, thermal fields exacerbate degradation or recombination, and electric fields require additional energy input while potentially accelerating corrosion[28,29]. In contrast, magnetic modulation technology presents unique advantages as a non-contact, green, and highly efficient strategy[30,31]. Its core mechanism lies in the efficient separation of photogenerated carriers through the directional manipulation of spin-polarized electrons[32–35]. Specifically, under an applied magnetic field (MF), the spin states of charge carriers are modulated. Conduction band (CB) electrons become spin-polarized and enhanced polarization, adopting spin orientations opposite to those of the valence band (VB) holes. Influenced by hyperfine interactions and spin-orbit coupling[36–38], a portion of excited electrons undergo spin-flip, while VB holes largely maintain their original orientation under high spin polarization. According to the Pauli exclusion principle, recombination between electron-hole pairs with opposite spins is significantly suppressed[39]. The MF acts as a continuous driving force, persistently modulating the spin states of electrons within the photoelectrode to maintain carrier separation, thereby providing an additional gain for photocurrent conversion. Furthermore, the MF can reduce material resistance via the negative magnetoresistance (MR) effect, enabling photogenerated charge carriers to transport with reduced resistance under identical illumination, consequently enhancing photoelectric conversion efficiency[40,41]. For example, Under an applied MF, $Mn^{2+}$-doped $CsPbBr_3$ perovskite nanosheets exhibited enhanced photocatalytic $CO_2$ reduction efficiency due to the synergistic effects of increased carrier spin polarization, prolonged carrier lifetime, and suppressed charge recombination[42]. In a PEC system, $ZnFe_2O_4$ exhibits ~150% photocurrent enhancement in PEC system via electron spin polarization. Following photoexcitation, which generates holes and electrons with opposite spins, spin randomization post-relaxation suppresses electron-hole recombination via the Pauli exclusion principle, while the MR effect provides more efficient charge transport pathways[43]. Despite the evident advantages of magnetic modulation, its practical application in PEC biosensing faces significant challenges. A key unresolved issue is the effective integration of the physical modulation benefits of magnetic fields with biorecognition elements to construct specific recognition sensing interfaces endowed with magnetically responsive regulation capabilities[44].

To address the aforementioned challenges, this study demonstrated a high-performance PEC biosensing platform for detecting protein kinase activity, using PKA as a model. We innovatively synthesized magnetic cobalt-doped titanium dioxide (Co-TiO$_2$) and integrated it with highly conductive MXenes to construct MXenes/Co-TiO$_2$ composite probes. Thes probes combine magnetic responsivity, high conductivity, broadband-spectral light-harvesting capability, and phosphorylation recognition functionality. Furthermore, the built-in electric field at the heterojunction synergistically enhances ion transport and electron-hole separation. As illustrated in Figure 1, Ti$^{4+}$ sites within the MXenes/Co-TiO$_2$ composite selectively recognize phosphate groups on the kinase phosphorylation substrate (kemptide) via coordination interactions[45]. Upon visible light excitation, photogenerated electrons from Co-TiO$_2$ rapidly transfer to the electrode surface through the conductive network of MXenes, generating a measurable photocurrent signal. The intensity of this signal is proportional to PKA activity. The core breakthrough of this work lies in the introduction of an external magnetic field. This field induces the formation of spin-mismatched photogenerated charge carriers, suppressing electron-hole recombination. Through a magneto-opto-electronic synergistic effect at the heterojunction interface, unprecedented efficiency in charge separation and photoelectric conversion was achieved. This study not only establishes a highly sensitive and reliable novel method for kinase activity analysis but also pioneers a transformative magneto-regulated strategy. This advancement propels the development of high-performance PEC biosensors and holds considerable potential for applications in early disease diagnosis and drug screening.

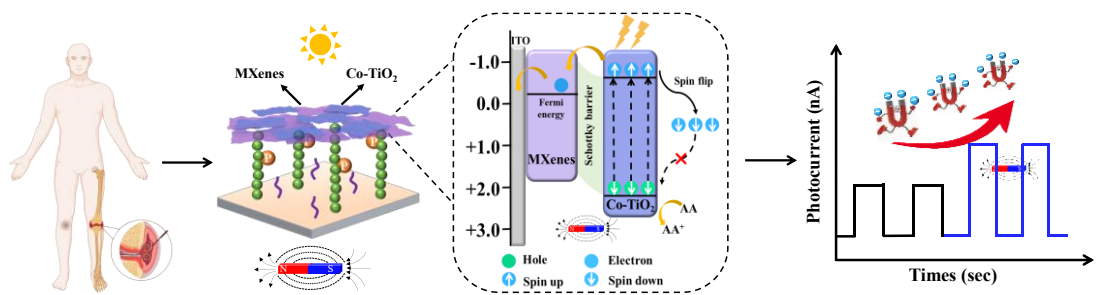

**Fig. 1** Schematic illustration of the mechanism of magnetic field-enhanced photoelectrochemical sensing technology for kinase activity detection

## Results

**Characterization of MXenes, Co-TiO$_2$ and MXenes/Co-TiO$_2$**

To characterize the morphological structure of the synthesized materials, the transmission electron microscopy (TEM) images of MXenes and Co-TiO$_2$ were acquired respectively. As shown in Figures 2a and 2b, the distinct edge contours of the nanosheets confirmed their successful synthesis, which is critical for constructing uniform and well-aligned nano-probes. Supplementary Fig. S1 shows the scanning electron microscopy (SEM) image of MXenes film, revealing a layered stacking structure. Supplementary Fig. S2 presents the atomic force microscopy (AFM) image of Co-TiO$_2$ nanosheet, with lateral dimensions of approximately 395.22 nm and a thickness of ~4.28 nm. Figure 2c exhibits the TEM image of the MXenes/Co-TiO$_2$ hybrid probes, which clearly demonstrates the tight interfacial integration between MXenes and Co-TiO$_2$ nanosheets. Figure 2d illustrates the suspensions of MXenes and Co-TiO$_2$, respectively, highlighting their excellent colloidal stability. The energy dispersive X-ray spectroscopy (EDS) mappings of the MXenes/Co-TiO$_2$ hybrid probes revealed the successful recombination of the material, as the Co, Ti, O, C, F, and Al elements were uniformly dispersed throughout the material, as shown in Figure 2e (1-6). The EDS mappings of MXenes and Co-TiO$_2$ are shown in Supplementary Fig. S3 and S4 respectively. The X-ray photoelectron spectroscopy (XPS) of MXenes powder (Supplementary Fig. S5) is consistent with previous reports, corroborating the successful synthesis of MXenes nanosheets[46,47]. The crystal structure of the synthesized materials was analyzed by X-ray diffraction (XRD). As shown in Figure 2f, the sharp and intense (002) peak for MXenes and the (020) peak for Co-TiO$_2$ indicate their successful synthesis, with the latter corresponding to periodic diffraction from the layered stacking structure[48,49]. The prominent peak in the MXenes/Co-TiO$_2$ composite nanocomposites suggests a highly ordered stacking configuration. Magnetic hysteresis loops of the Co-TiO$_2$ dispersion confirm its superior magnetic properties, as shown in Figure 2g. The optical absorption characteristics of Co-TiO$_2$ were probed by UV-vis spectroscopy, revealing its strong visible-light absorption shown in Figure 2h. The inset in Figure 2h displays the Tauc plot derived from the Kubelka-Munk function, yielding a bandgap of 2.87 eV for Co-TiO$_2$. Furthermore, the energy band structure and positions of Co-TiO$_2$ were analyzed via Mott-Schottky (MS) plots. As shown in Figure 2i, Co-TiO$_2$ exhibited a positive slope in its MS curve, confirming its n-type semiconductor behavior. The flat-band potential ($E_{fb}$) was determined to be -0.47 eV by extrapolating the tangent of the curve to the x-axis. Based on the equations provided in Supplementary Note 1, the conduction band potential ($E_{CB}$) and valence band potential ($E_{VB}$) of Co-TiO$_2$ were calculated as -0.67 V and 2.2 eV, respectively. In the MXenes/Co-TiO$_2$ heterojunction, the lower Fermi level of MXenes induced a pronounced Schottky-derived built-in electric field at the interface[50]. This field efficiently drove spatial separation and directional migration of photogenerated carriers, significantly suppressing recombination and enhancing electron utilization efficiency.

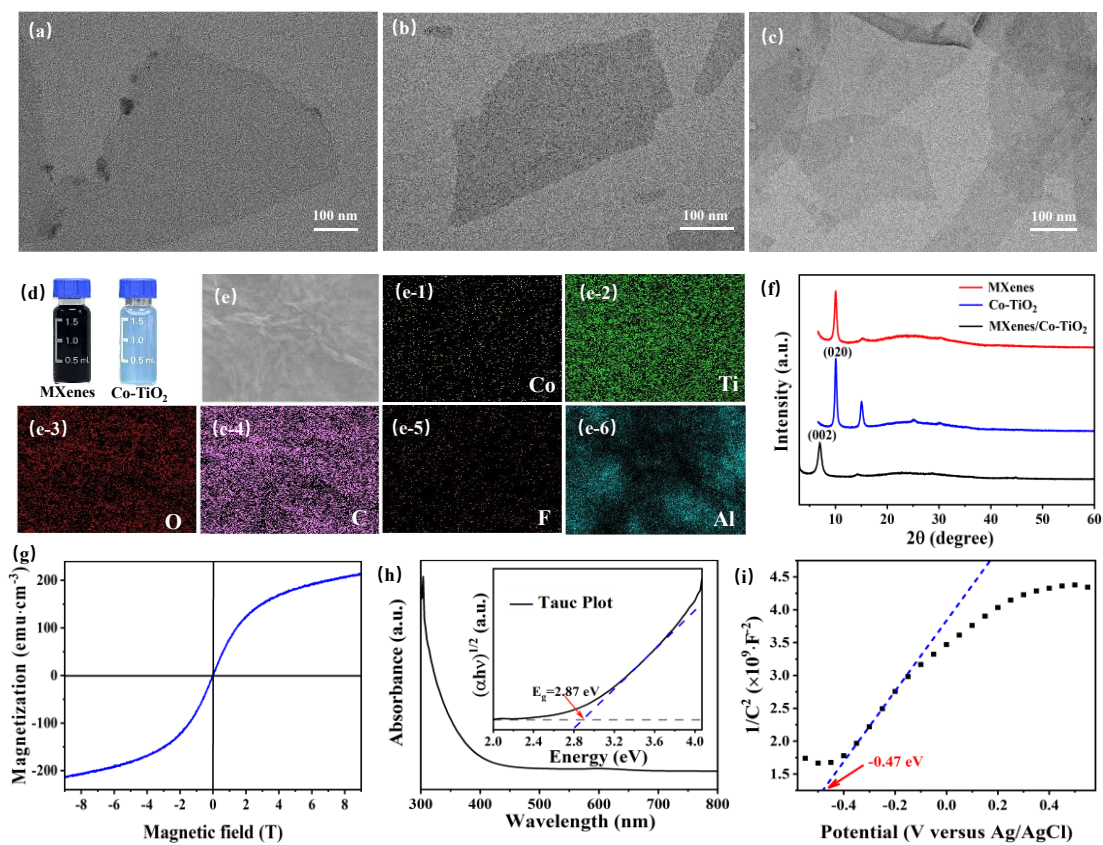

**Fig. 2 Material characterization and performance analysis.** TEM images of (a) MXenes, (b) Co-TiO$_2$, and (c) MXenes/Co-TiO$_2$ composite probes. (d) Pictures of MXenes and Co-TiO$_2$ suspensions. (e) SEM image of MXenes/Co-TiO$_2$ composite probes along with the corresponding EDS elemental mappings (1-6), showing the elemental composition of Co:Ti:O:C:F:Al = 0.42:4.16:17.76:69.21:0.5:7.94. (f) XRD pattern of the samples. (g) Hysteresis loop of Co-TiO$_2$ suspensions. (h) UV-Vis absorption spectrum of Co-TiO$_2$ suspensions, with the inset Tauc plot indicating a band gap of 2.8 eV. (i) Mott-Schottky curve of Co-TiO$_2$, with the negative slope of the tangent line indicating n-type semiconductor behavior and the flat-band potential at -0.40 eV.

**Biosensor fabrication and PEC behaviors**

The prepared MXenes/Co-TiO$_2$ composite probes were used to fabricate the biosensor. The assembly procedure of the biosensor is illustrated in Supplementary Fig. S6. First, ITO conductive glass was ultrasonically cleaned in acetone, isopropanol, and ethanol (15 min each) for surface pretreatment, followed by drying under nitrogen gas to serve as the substrate. Subsequently, a multilayered modification was constructed via stepwise deposition on a defined ITO area (0.5 cm²). Firstly, 20 μL of 0.5% (v/v) chitosan solution was drop-cast and air-dried at room temperature. Secondly, 20 μL of 2.5% (v/v) glutaraldehyde solution was applied and reacted for 1 hour to enable covalent crosslinking between its bifunctional aldehyde groups and the amine groups of chitosan, followed by immobilization of 20 μL of kemptide (50-500 mM) for 6-8 hours in the dark to ensure stable conjugation. Thirdly, Non-specific binding sites were blocked by treating 1 mM 6-aminohexanoic acid for 2 hours. The electrode was then incubated with 20 μL of PBS solution (i.e., 50 mM Tris-HCl buffer containing 20 mM $MgCl_2$, pH 7.4) containing gradient concentrations of PKA and adenosine triphosphate (ATP) for 2-3 hours to facilitate phosphorylation. Finally, 20 μL of the MXene/Co-$TiO_2$ hybrid dispersion was drop-cast to form the signal-responsive interface, where $Ti^{4+}$ in MXenes and Co-$TiO_2$ synergistically coordinated to capture phosphate groups. After solvent evaporation, the PEC biosensor was fabricated and stored at 4°C for subsequent use.

The PEC behaviors of the modified biosenso were further investigated using photocurrent transient measurements. As shown in Supplementary Fig. S7, the photocurrent responses of the modified ITO electrode in 0.1 M PBS containing 0.1 M AA under visible light irradiation. Bare ITO electrodes and chitosan-modified ITO electrodes exhibited negligible photo-response under these conditions. Similarly, no detectable photocurrent was observed when the chitosan/ITO electrode was functionalized with kemptide and subsequently phosphorylated by PKA. In contrast, a remarkable enhancement in photocurrent signal was achieved after immobilizing MXenes/Co-$TiO_2$ hybrid probes onto the kemptide-modified electrode. This phenomenon can be attributed to the efficient transfer of photogenerated electrons from MXenes/Co-$TiO_2$ to the electrode surface under visible-light excitation, thereby generating a measurable photocurrent.

**Magnetic field enhanced Photoelectrochemical strategy**

In order to enhance the PEC performance of the prepared biosensor, a magnetic-regulated strategy was introduced. The schematic diagram of the device for magnetic field enhanced PEC signals is shown in Figure 3a. The photocurrent response was investigated under varying relative positions of the magnet to the sensing electrode. As shown in Figure 3b, the photocurrent signal was significantly enhanced at Position (3), while no enhancement was observed at other positions (1and 2) due to insufficient magnetic flux lines perpendicular to the sensing electrode. This observation highlights the critical role of magnetic flux line orientation in the magnetization of the sensing electrode. Furthermore, Figure 3c compares the photocurrent intensity variations with and without magnetic field application. The photocurrent variation intensity ($\Delta I$) was calculated using the formula $\Delta I = I_{light\ on} - I_{light\ off}$, where $I_{light\ on}$ and $I_{light\ off}$ represent the photocurrent signal intensities under illumination and dark conditions, respectively. The calculated results demonstrate a 58% enhancement in photocurrent variation intensity under MF application. These indicated that the magnetic field plays a crucial role in regulating PEC performance. Among the contributing factors, magnetic field-induced spin polarization emerges as a key physical parameter with broad application potential, as it significantly influences the PEC signals. Based on first-principles analysis of the Co-TiO$_2$ samples (Figures 3d-i, with relevant parameters provided in Supplementary Note 2), the spin-polarized electron density distribution revealed distinct spatial characteristics of spin distribution. In the absence of a MF, Co doping introduced isolated electronic states within the bandgap, which were confined to lower energy levels (<1 eV) (Figure 3d). However, upon applying an external MF, a pronounced spin splitting occurred between spin-up and spin-down states, leading to the emergence of high-energy, strongly spin-polarized electronic states near the Fermi level, while the isolated mid-gap states vanished (Figure 3e). 3D charge density difference and planar plots of Co-TiO$_2$ further demonstrated that spin-polarized charges were mainly concentrated around the Co and O atoms. Notably, in the absence of a magnetic field, the electron spins of Co and its coordinated O atoms were aligned in the same direction (Figures 3f and 3h). In contrast, under a magnetic field, partial spin inversion was observed in some Co atoms and their coordinated O atoms (Figures 3g and 3i). These magnetic field-induced changes in spin configuration effectively suppressed electron-hole recombination, thereby significantly enhancing the material's photoelectric conversion efficiency.

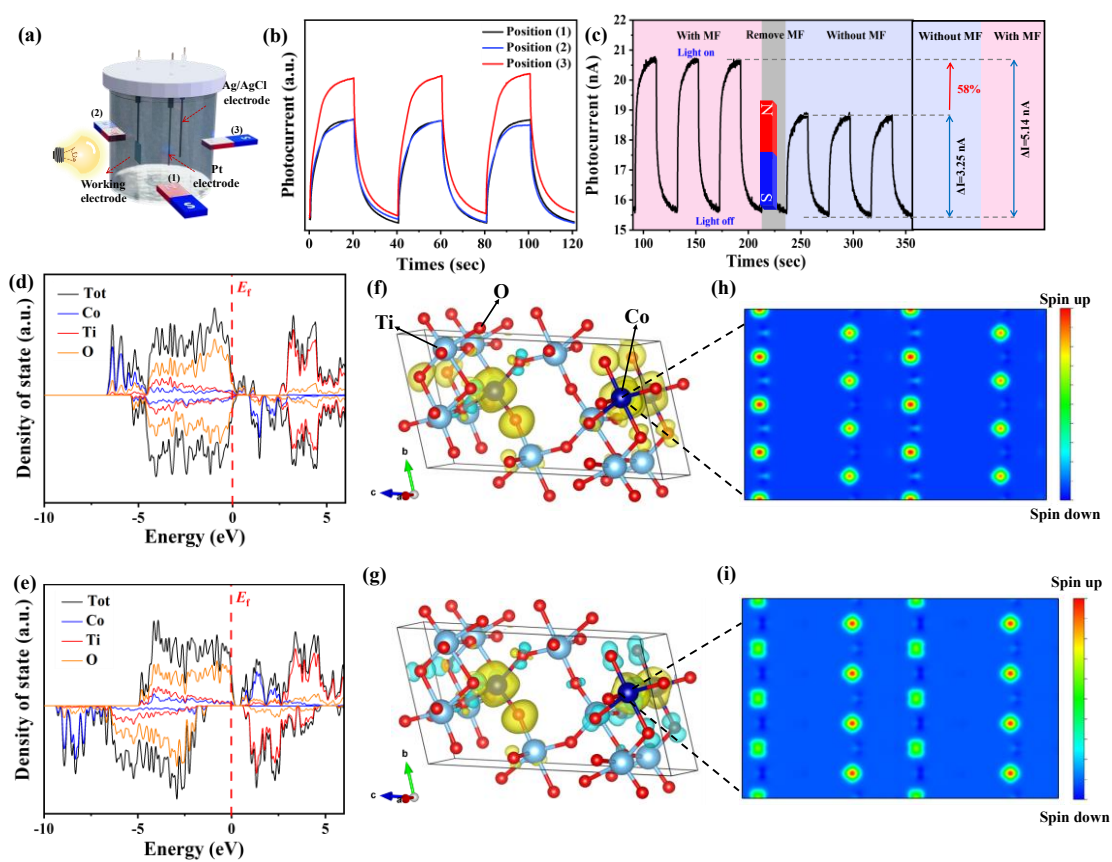

**Fig. 3 Photochemical performance and spin properties under MF conditions.** (a) Schematic diagram of the PEC device with different magnet arrangements relative to the working electrode and (b) current-time (I-t) response curves corresponding to different magnet positions, where the configuration at position (3) shows a significant influence on photocurrent variation. (c) I-t curves of the system under MF conditions at position (3) compared with those without the MF. The predicted density of states (PDOS) of different atoms of Co-TiO$_2$ without and with magnetic field (d and e). The 3D charge density difference of Co-TiO$_2$ without and with magnetic field, respectively (f and g), the yellow and cyan regions represent electron accumulation and depletion, respectively. The corresponding 2D spin density maps in the (010) direction (h and i).

**Magnetic field suppressing photo-induced carrier recombination**

To investigate the magnetic field-modulated separation characteristics of photoinduced charge carriers, MR profiles, EIS, MS plots, OCPD technique and PL spectra were systematically analyzed, as shown in Figure 4. MR, defined as the relative resistance change induced by spin-polarized electrons under the MF, was calculated using the formula MR% = [R(H)-R(0)]/R(0), where R(H) and R(0) represent the material's resistance with and without the MF, respectively. The MR of the Co-TiO$_2$ sample was recorded at room temperature under MF ranging from -3 to 3 T, as shown in Figure 4a. A negative MR effect of approximately -5.3% was observed at ±3 T, attributed to magnetic field-induced electron spin polarization. This negative MR indicates enhanced conductivity under MF, confirming that the field effectively promotes the transfer of photogenerated electrons to the electrode surface. EIS further validated the MF's improvement of charge transfer kinetic, as shown in Figure 4b. Under illumination, the Nyquist plot of MXene/Co-TiO$_2$ exhibited a significant reduction in charge transfer resistance ($R_{et}$) upon MF application, demonstrating that the field optimizes carrier transport pathways and lowers the energy barrier at the electrode-electrolyte interface. Combined with MS analysis in Figure 4c, the negative slope of the MS curve confirms the n-type semiconductor behavior of the material, with photogenerated electrons as the dominant charge carriers. According to the MS relationship, slop=2($\varepsilon\varepsilon_0 A^2 eN_D$)$^{-1}$, where $\varepsilon$ and $\varepsilon_0$ are dielectric constants, $A$ is the electrode area, and $N_D$ is the dopant density in the semiconductor, which can be assigned to the charge carrier density[51]. the smaller MS slope under MF indicated a higher photogenerated carrier concentration in MXenes/Co-TiO$_2$. Notably, the flat-band potential ($E_{fb}$) remained constant across varying illumination/magnetic conditions, confirming that the MF enhances conductivity by optimizing carrier distribution rather than altering the intrinsic semiconductor properties. Furthermore, the OCPD technique was employed to evaluate the positive effect of the MF on promoting the separation of photogenerated charge carriers. Under visible light irradiation, the photoelectrode absorbed photon energy, resulting in a decrease in open-circuit voltage ($V_{oc}$). Once the light reached a steady-state condition and the illumination was turned off, $V_{oc}$ gradually recovered to its initial value over time. The lifetime of the photogenerated charge carriers ($\tau_n$) was calculated using the equation $\tau_n=-(k_BT/e)(dV_{oc}/dt)^{-1}$, where $k_B$ was the Boltzmann constant, T was the temperature, and e was the elementary charge[52]. As shown in Figure 4d, under the influence of a MF, the $\tau_n$ value of MXenes/Co-TiO$_2$ was significantly prolonged compared to the condition without a magnetic field. This indicated that the MF increased the number of spin-polarized photogenerated charge carriers, effectively suppressed carrier recombination, and thus extended the carrier lifetime. Additionally, PL spectra further elucidated the MF's regulatory effect on charge recombination dynamics. Under 532 nm laser excitation, the PL intensity of Co-TiO$_2$ at 650 nm decreased by 1.6% under MF (Supplementary Fig. S8), while the MXenes/Co-TiO$_2$ composites exhibited a more pronounced 12% reduction, as shown in Figure 4e. This suppression of PL intensity, coupled with prolonged carrier lifetimes in the MXenes/Co-TiO$_2$ system under MF, confirms superior synergistic enhancement of charge separation and transport efficiency.

Figure 4f schematically illustrates the mechanism underlying the magnetic field-enhanced photocurrent signal in the sensing platform. Owing to the modification with MXenes and Co doping in TiO$_2$, the MXenes/Co-TiO$_2$ hybrid probes exhibit enhanced broad-spectrum light absorption, particularly with significantly improved absorption efficiency in the visible region. Upon visible-light irradiation, electron-hole pairs are generated in the probe. Given the lower Fermi level of MXenes compared to Co-TiO$_2$, photogenerated electrons rapidly transfer from Co-TiO$_2$ to the MXenes surface and further migrate to the counter electrode, generating a photocurrent in the external circuit. Concurrently, the accumulation of holes and electrons at the MXenes and Co-TiO$_2$ interface establishes a built-in electric field. This field induces band bending in Co-TiO$_2$ and forms an energy barrier, which provides a unidirectional pathway for electron transfer and charge separation, effectively suppressing carrier recombination. In here, AA acts as an electron donor in this system, efficiently scavenging holes, thereby minimizing electron-hole recombination and amplify photocurrent intensity. Meanwhile, an external MF generated by permanent magnets is applied to manipulate spin-polarized electrons, promoting the efficient separation of photoexcited electron-hole pairs and achieving unprecedented photoconversion efficiency. Specifically, Under the influence of a magnetic field, the holes and excited state electrons produced by spin-polarized electrons after optical excitation have opposite directions. After relaxation of the electrons in the excited state, their spin direction tends to become random, which suppresses the recombination of electrons and holes based on the Pauli exclusion principle, thereby enhancing the utilization of charge carriers; Furthermore, the magneto resistance effect of Co-TiO$_2$ under MF optimizes carrier mobility and charge separation efficiency, synergistically enhancing photocurrent signals and enabling ultrasensitive detection of PKA.

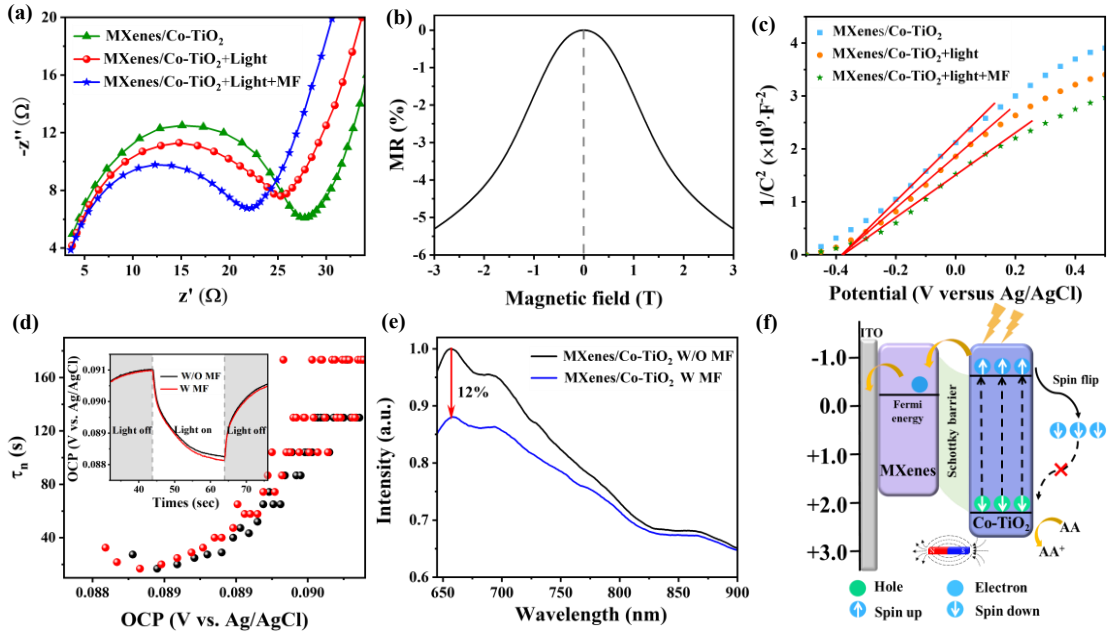

**Fig.4 Experimental verification of carrier recombination with MF.** (a) EIS changes of the MXenes/Co-TiO$_2$ hybrid probe under dark, light, and light-magnetic synergistic conditions. (b) Magnetoresistance variation rate of Co-TiO$_2$ dispersion at room temperature. (c) MS plots of the MXenes/Co-TiO$_2$ hybrid probe under dark, light, and light-magnetic synergistic conditions. (d) Average carrier lifetime($\tau_n$) calculated through OCPD measurement for MXenes/Co-TiO$_2$. (e) PL spectra of MXenes/Co-TiO$_2$ hybrid probe with MF and without MF conditions. (f) Electron transfer mechanism of the MXenes/Co-TiO$_2$ heterojunction under MF conditions.

**Optimization of the fabrication conditions**

To achieve optimal performance of the PEC biosensor, the preparation parameters of the sensor were optimized under identical external magnetic field conditions. In this study, kemptide, a specific peptide substrate which can be phosphorylated by PKA was optimized. As shown in Supplementary Fig. S9a, the photocurrent variation intensity of the modified electrode increased with elevated kemptide concentrations and reached a maximum at 500 μM. Therefore, the optimal kemptide concentration was determined to be 500 μM. Phosphorylation of kemptide typically requires the presence of ATP, which not only serves as a phosphate group donor but also supplies essential energy for the catalytic activity of the kinase. As depicted in Supplementary Fig. S9b, the photocurrent signal of the electrode increased with higher ATP concentrations, achieving a maximum at 120 μM, indicating that this concentration was sufficient to support the PKA-catalyzed phosphorylation reaction. Thus, the optimal ATP concentration was identified as 120 μM. Furthermore, phosphorylation time is a critical parameter in electrode preparation. As illustrated in Supplementary Fig. S9c, the photocurrent signal gradually increased with prolonged incubation time and plateaued at 80 minutes, suggesting near-complete phosphorylation of kemptide. Consequently, the optimal phosphorylation time was established as 80 minutes.

**PEC measurements of PKA activity**

Under optimized experimental conditions, the kinase activity of PKA was evaluated using varying concentrations. As shown in Figure 5a, the photocurrent exhibited a concentration-dependent increase with PKA activity, reaching a plateau at 80 U/mL. The inset of Figure 5b demonstrates a linear correlation between photocurrent intensity and PKA concentration within the range of 0.005 to 1 U/mL, described by the equation $y = 4.93 + 62.36x$ (where $y$ represents photocurrent intensity and $x$ represents PKA concentration), with a correlation coefficient of $R^2 = 0.998$. The detection limit (LOD) for PKA was determined to be 0.00016 U/mL (S/N = 3), which was calculated by the formula of LOD = (3σ/k), where σ and k represent standard deviation of blank value and slope of the linear regression equation, respectively[53]. In contrast, the LOD was calculated to be 0.00027 U/mL (S/N = 3) without the addition of MF, and maintaining the same correlation coefficient ($R^2 = 0.998$), as shown in Supplementary Fig. S10. As a result, the newly developed platform demonstrated a 68.75% enhancement in detection sensitivity compared to the same probe-modified biosensor without the magnetic field. This methodology demonstrates significantly enhanced detection performance compared to previously reported approaches. Relative to other research efforts in the last 5 years, the LOD of the prepared biosensors had increased by at least 2–3 orders of magnitude. (see Table S1 in Supporting Information).

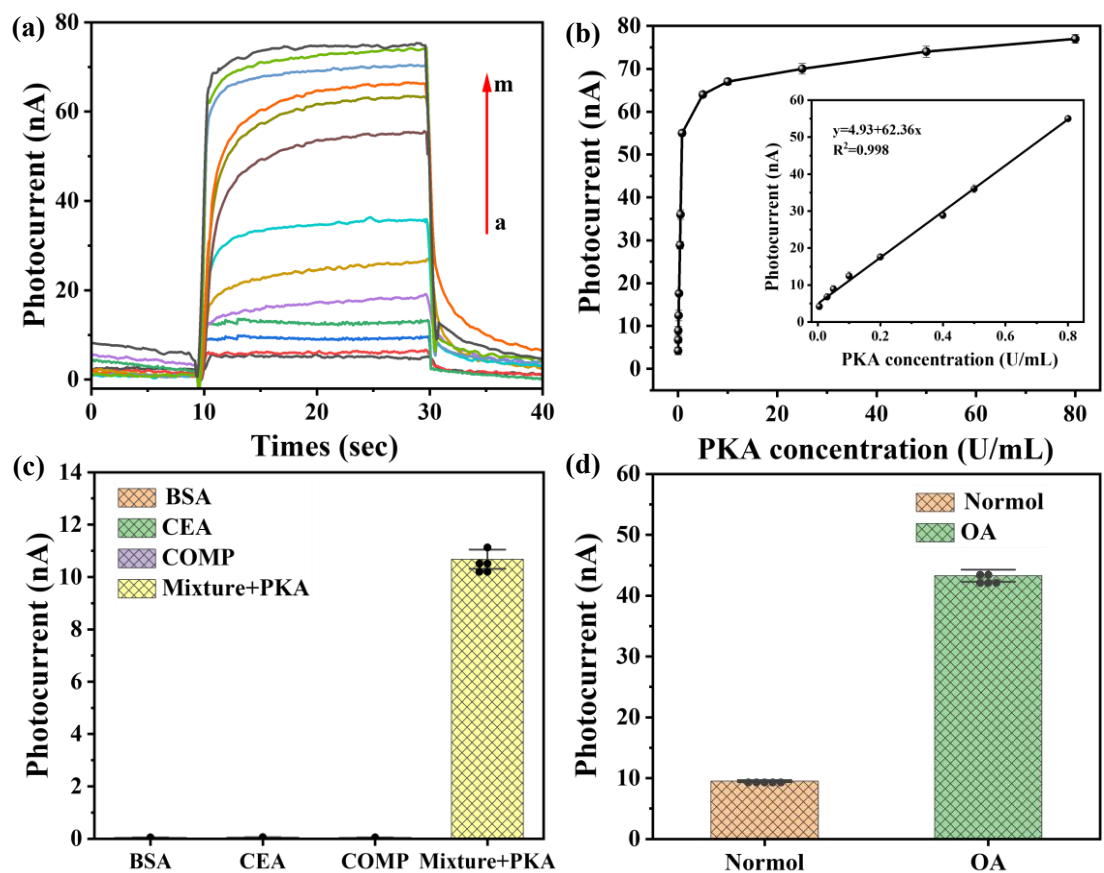

**Fig. 5 PEC sensing for PKA activity detection.** (a) Photocurrent response curves at different PKA concentrations: a) 0.005 U/mL, b) 0.03 U/mL, c) 0.05 U/mL, d) 0.1 U/mL, e) 0.2 U/mL, f) 0.4 U/mL, g) 0.5 U/mL, h) 0.8 U/mL, i) 5 U/mL, g) 10 U/mL, k) 25 U/mL, l) 50 U/mL, m) 80 U/mL. (b) Photocurrent variation intensity at different PKA concentrations ranging from 0.005 to 80 U/mL under magnetic field conditions. The inset shows the linear relationship between current intensity and PKA concentration. (c) Selectivity and interference detection of the PEC biosensor for PKA: Selectivity response at 10 U/mL concentrations of interfering substances (BSA, CEA, COMP) and in the presence of 0.5 U/mL PKA. (d) PKA activity detection in knee joint synovial fluid from normal individuals and osteoarthritis patients using the PEC biosensing platform.

To validate the excellent selectivity of the fabricated biosensor, three proteins-BSA, CEA, and COMP, which cannot recognize the kemptide substrate peptide-were selected as model interferents. As illustrated in Figure 5c, the current change upon exposure of the sensor to PKA was obviously higher than those of the interference groups, demonstrating the specific recognition capability of the developed biosensor toward PKA. The clinical applicability of the photoelectrochemical sensor was evaluated by analyzing 5000-fold diluted clinical samples. As shown in Figure 5d, the photocurrent changes intensity of the sensor fabricated with synovial fluid samples from osteoarthritis patients was markedly higher than that of the healthy control group, with minimal detection errors. These findings are consistent with the results obtained using a commercial ELISA kit (Supplementary Fig. S11). The stability of the biosensor was verified through eight consecutive cyclic tests (Supplementary Fig. S12), which confirmed excellent detection reproducibility. Experimental results demonstrate that the proposed electrochemical sensing method enables highly sensitive detection of PKA activity changes in synovial fluid associated with joint inflammation, providing a novel and highly reliable analytical tool for early diagnosis and dynamic monitoring of joint inflammatory conditions.

## Discussion

This study successfully developed a magnetic field-enhanced MXenes/Co-TiO$_2$ heterojunction photoelectrochemical biosensing platform for ultrasensitive detection of PKA activity. The integration of MXenes and doping of Co not only broadened MXenes/Co-TiO$_2$ visible-light absorption range but also enhanced their conductivity. Furthermore, systematic investigations revealed that an external magnetic field significantly suppressed photogenerated electron-hole recombination of MXenes/Co-TiO$_2$ probes by inducing carrier spin state mismatch. The proposed strategy enhanced the photocurrent signal variation intensity by 58%. Correspondingly, the developed platform showed a 68.75% improvement in detection sensitivity compared to the same probe-modified biosensor without the magnetic field. Moreover, the fabricated biosensor demonstrated outstanding performance with an ultra-low detection limit of 0.00016 U/mL and a broad linear response (0.005-80 U/mL). Thus, the platform exhibited remarkable specificity and operational stability, with clinical validation using synovial fluid from osteoarthritis patients showing excellent correlation with commercial ELISA kit results. Innovatively combining magnetic modulation strategies with two-dimensional nanomaterial heterojunctions, this work provides a universal solution to overcome intrinsic carrier recombination limitations in photoelectric systems and a promising tool for early disease diagnostics and pharmaceutical screening applications.

## Materials and methods

### Materials and Reagents

The indium tin oxide (ITO) conductive glass was obtained from Huanan Xiangcheng Technology Co., Ltd. (Shenzhen, China). Cysteine-terminated kemptide (CLRRASLG) was obtained from GL Biochem (Shanghai, China). Glutaraldehyde (25% in $H_2O$) and 6-aminohexanoic (99%) were both obtained from Macklin Biochemical Co., Ltd. (Shanghai, China), PKA (catalytic subunit from bovine heart) was obtained from Sigma-Aldrich corp., Adenosin triphosphate (ATP) was obtained from Dingguo Biological Products Company (China), Tris-HCl buffer (PH 7.4) was obtained from Shanghai Macklin Biochemical Technology Co., Ltd., Bovine serum albumin (BSA) was obtained from Sangon Biotech (Shanghai, China) Co., Ltd. COMP, Carcinoembryonic antiden (CEA) was obtained from BBI life sciences corp. (Shanghai, China). Polyclonal antigen to cartilage oligomeric matrix protein (COMP) was obtained from cloud-clone corp. (USA). Other regents of analytical grade were provided from Beijing Chemical Company (China).

### Apparatus and Characterization

The electrochemical measurements were performed using a CHI802B electrochemical workstation (CH Instruments, Shanghai, China) with a conventional three-electrode system: a modified electrode as the working electrode, a platinum wire as the counter electrode, and an Ag/AgCl electrode (saturated KCl solution) as the reference electrode. Photocurrent responses and Open Circuit Optical Voltage Attenuation (OCPD) test were recorded in 0.1 M phosphate-buffered saline (PBS, pH 7.4) containing 0.1 M ascorbic acid (AA) under visible light irradiation ($\lambda$ = 420 nm, xenon lamp with a light intensity of 190 mW/cm²). Electrochemical impedance spectroscopy (EIS) was conducted in a solution containing 5 mM $K_3[Fe(CN)_6]/K_4[Fe(CN)_6]$ and 0.1 M KCl, with a frequency range of 0.1 Hz to 100 kHz. Mott-Schottky analysis was carried out in 0.1 M $Na_2SO_4$ solution at an AC frequency of 1000 Hz. The morphology and microstructure of the samples were investigated by scanning electron microscopy (SEM, SU8010, Hitachi, Japan) with energy dispersive X-ray spectroscopy (EDS) analysis, transmission electron microscopy (TEM, Tecnai G2 F30, FEI, USA), and atomic force microscopy (AFM, Dimension Icon, Bruker, Germany). Crystal structures were analyzed via X-ray diffraction (XRD, Bruker D8-Advance) with Cu K$\alpha$ radiation ($\lambda$=1.5418 Å). Surface elemental composition and chemical states were determined by X-ray photoelectron spectroscopy (XPS, PHI QUANTERA-II SXM, ULVAC-PHI, Japan). Hysteresis loop and magnetoresistance (MR) changes of Co-$TiO_2$ dispersions were measured using a Physical Property Measurement System (PPMS-9T, Quantum Design, USA). UV-Vis absorption spectra were acquired using a microplate reader (Varioskan Flash). Photoluminescence (PL) spectra were obtained with a confocal Raman microscope (alpha300 R, WITec, Germany) equipped with a 532 nm excitation laser. An external MF is applied using a rectangular permanent magnet (30 × 10 × 3 mm) with a magnetic induction intensity of 100 mT.

**Preparation of MXenes, Co-$TiO_2$ and MXenes/Co-$TiO_2$**

**Synthesis of Ti$_3$C$_2$ MXenes suspensions:** 0.8 g of LiF was first added to 10 mL of 9 M HCl and gently stirred until fully dissolved, Then, 0.5 g of Ti$_3$AlC$_2$ powder was added, and the mixture was stirred continuously at 35 °C for 24 h to carry out the etching reaction. Afterward, the product was washed by repeated centrifugation until the supernatant reached pH of 6. The resulting precipitate was dispersed in deionized water and centrifuged at 3500 rpm for 60 min to remove unexfoliated components. The sediment was then ultrasonicated under nitrogen for 1 h and centrifuged again under the same conditions. The final supernatant, rich in MXenes nanosheets, was collected and stored at 4 °C for further use.

**Synthesis of Co-TiO$_2$ suspensions:** To prepare Co-TiO$_2$ suspensions, we followed a four-stage approach, similar to previously reported but with some changes[54,55]. Firstly, TiO$_2$ (0.25 mol, 20 g), CoO (0.03 mol, 2.25 g), K$_2$CO$_3$ (0.06 mol, 5.94 g) and Li$_2$CO$_3$ (0.01 mol, 0.67 g) were mixed according to the stoichiometric ratio, and the compound was K$_{0.8}$Ti$_{(5.2-y)/3}$Li$_{(0.8-2y)/3}$Co$_y$O$_4$ after grinding two times annealing at 1000°C for 5 h and 20 h respectively. Secondly, the precursor was reacted with 200 mL HCl (1M) solution for 4 days, Li$^+$/K$^+$ was replaced by H$^+$, and H$_{(3.2-2y)/3}$Ti$_{(5.2-y)/3}$Co$_y$O$_4$ was obtained after washing and drying. Thirdly, the protonation product is soaked in tetrabutylammonium hydroxide (TBAOH) at a 1:1 molar ratio for 5 hours to realize the exchange of H$^+$ and TBA$^+$ to generate TBA$_z$H$_{(3.2-2y)/3-z}$Ti$_{(5.2-y)/3}$Co$_y$O$_4$. Fourth, the intercalation compound was stripped by mechanical shaking for 48 h, and finally a stable suspension of Co-TiO$_2$ was obtained.

**Preparation of MXenes/Co-TiO$_2$ signal probes:** The MXenes suspensions (1.48 mg/mL) and Co-TiO$_2$ suspensions (1 mg/mL) were mixed at a volume ratio of 10:1, thoroughly vortex-mixed, then allowed to stand for several hours, and were subsequently stored at 4°C for further use.

**Acknowledgements**

This work was supported by the National Natural Science Foundation of China (No. 22304013, No. 22174084), Beijing Natural Science Foundation (L251023) and Financial Program of BJAST (No. 25CB003-05).

**Author details**

[1]Beijing Engineering Research Center of Mixed Reality and Advanced Display, School of Optics and Photonics, Beijing Institute of Technology, Beijing, 100081, China.

[2]Institute of Analysis and Testing, Beijing Academy of Science and Technology (Beijing Center for Physical and Chemical Analysis), Beijing, 100089, China.

[3]Department of Chemistry, Kay Lab of Bioorganic Phosphorus Chemistry and Chemical Biology of Ministry of Education, Beijing Key Laboratory for Analytical Methods and Instrumentation, Tsinghua University, Beijing, 100084, China.

[4]Department of Sports Medicine Peking University Third Hospital, Institute of Sports Medicine of Peking University, Beijing Key Laboratory of Sports Injuries, Engineering Research Center of Sports Trauma Treatment Technology and Devices Ministry of Education, Beijing, 100191, China.

[5]School of Integrated Circuits and Electronics, Beijing Institute of Technology, Beijing, 100081 China.

**Author contributions**

Tao Wang conducted the experiment, numerical simulations and authored the manuscript, Yunhe An, Yunyun Dai, Nan Yang, Chaojie Yang, Xinran Zhou and Yingshan Ma performed the characterization of the material, Hongjie Huang and Yucheng Zhu provided the extraction and processing of human synovial samples, Yongrui Li proofread the manuscript. Nan Zhang, Yang Liu, Lingling Huang and Zhiyong Yan conceived the concept, supervised this project, and revised the manuscript. All authors participated in critical discussion of the final draft.

**Conflict of interest**

The authors declare no competing interests.

**References**

1. Shi, J., Li, P., Kim, S. & Tian, B. Implantable bioelectronic devices for photoelectrochemical and electrochemical modulation of cells and tissues. *Nat. Rev. Bioeng.* **3**, 485–504 (2025).
2. Li, L. Recent advances in photoelectrochemical sensors for detection of ions in water. *Chin. Chem. Lett.* **34**, 107904 (2023).

3. Kaur, M., Kumar, P. & Ghotra, H. S. A review on advances in photoelectrochemical (PEC-type) photodetectors: A trending thrust research area. *Int. J. Hydrog. Energy* **49**, 1095–1112 (2024).
4. Wang, D. *et al.* Observation of polarity-switchable photoconductivity in III-nitride/MoS$_x$ core-shell nanowires. *Light Sci. Appl.* **11**, 227 (2022).
5. Shu, J. & Tang, D. Recent Advances in Photoelectrochemical Sensing: From Engineered Photoactive Materials to Sensing Devices and Detection Modes. *Anal. Chem.* **92**, 363–377 (2020).
6. Tu, W., Wang, Z. & Dai, Z. Selective photoelectrochemical architectures for biosensing: Design, mechanism and responsibility. *TrAC Trends Anal. Chem.* **105**, 470–483 (2018).
7. Lv, M. *et al.* A novel electrochemical biosensor based on MIL-101-NH$_2$ (Cr) combining target-responsive releasing and self-catalysis strategy for p53 detection. *Biosens. Bioelectron.* **214**, 114518 (2022).
8. Gao, R.-T. *et al.* Single-atomic-site platinum steers photogenerated charge carrier lifetime of hematite nanoflakes for photoelectrochemical water splitting. *Nat. Commun.* **14**, 2640 (2023).
9. Fu, G. *et al.* Construction of Thiadiazole-Bridged sp$^2$-Carbon-Conjugated Covalent Organic Frameworks with Diminished Excitation Binding Energy Toward Superior Photocatalysis. *J. Am. Chem. Soc.* **146**, 1318–1325 (2024).
10. Mo, F. *et al.* Recent Advances in Photoelectroanalysis: Carbon-Containing Materials for Enhanced Sensing Performance. *Adv. Funct. Mater.* 2504679 (2025).
11. Zhang, W. J. *et al.* Electronic and mechanical properties of monocrystalline silicon doped with trace content of N or P: A first-principles study. *Solid State Sci.* **120**, 106723 (2021).
12. Jing, M. *et al.* Coral-like B-doped g-C$_3$N$_4$ with enhanced molecular dipole to boost photocatalysis-self-Fenton removal of persistent organic pollutants. *J. Hazard. Mater.* **449**, 131017 (2023).
13. Arumugam, M., Koutavarapu, R., Seralathan, K.-K., Praserthdam, S. & Praserthdam, P. Noble metals (Pd, Ag, Pt, and Au) doped bismuth oxybromide photocatalysts for improved visible light-driven catalytic activity for the degradation of phenol. *Chemosphere* **324**, 138368 (2023).
14. Zhang, F. Room temperature photocatalytic deposition of Au nanoparticles on SnS$_2$ nanoplates for enhanced photocatalysis. *Powder Technol.* **383**, 371–380 (2021).
15. Han, T. *et al.* Anion-exchange-mediated internal electric field for boosting photogenerated carrier separation and utilization. *Nat. Commun.* **12**, 4952 (2021).
16. Gherabli, R., Indukuri, S. R. K. C., Zektzer, R., Frydendahl, C. & Levy, U. MoSe$_2$/WS$_2$ heterojunction photodiode integrated with a silicon nitride waveguide for near infrared light detection with high responsivity. *Light Sci. Appl.* **12**, 60 (2023).
17. Ding, N. *et al.* Highly DUV to NIR-II responsive broadband quantum dots heterojunction photodetectors by integrating quantum cutting luminescent concentrators. *Light Sci. Appl.* **13**, 289 (2024).
18. Tan, F. *et al.* Physisorption-assistant optoelectronic synaptic transistors based on Ta$_2$NiSe$_5$/SnS$_2$ heterojunction from ultraviolet to near-infrared. *Light Sci. Appl.* **14**, 122 (2025).
19. Wang, G. *et al.* Advancements in heterojunction, cocatalyst, defect and morphology engineering of semiconductor oxide photocatalysts. *J. Materiomics* **10**, 315–338 (2024).

20. Xue, S., Gao, Y., Wang, B. & Zhi, L. Effects of external physical fields on electrocatalysis. *Chem Catal.* **3**, 100762 (2023).
21. Wang, Z., Li, Y., Wu, C. & Tsang, S. C. E. Electric-/magnetic-field-assisted photocatalysis: Mechanisms and design strategies. *Joule* **6**, 1798–1825 (2022).
22. Wu, J., Qin, N. & Bao, D. Effective enhancement of piezocatalytic activity of $BaTiO_3$ nanowires under ultrasonic vibration. *Nano Energy* **45**, 44–51 (2018).
23. Hong, D. *et al.* High Piezo-photocatalytic Efficiency of CuS/ZnO Nanowires Using Both Solar and Mechanical Energy for Degrading Organic Dye. *ACS Appl. Mater. Interfaces* **8**, 21302–21314 (2016).
24. Chen, J. *et al.* Synergistic effect of photocatalysis and pyrocatalysis of pyroelectric $ZnSnO_3$ nanoparticles for dye degradation. *Ceram. Int.* **46**, 9786–9793 (2020).
25. Dai, B. *et al.* Construction of Infrared-Light-Responsive Photoinduced Carriers Driver for Enhanced Photocatalytic Hydrogen Evolution. *Adv. Mater.* **32**, 1906361 (2020).
26. Tank, C. M. *et al.* Electric field enhanced photocatalytic properties of $TiO_2$ nanoparticles immobilized in porous silicon template. *Solid State Sci.* **13**, 1500–1504 (2011).
27. Guo, H. *et al.* $CO_2$ Capture on *h*-BN Sheet with High Selectivity Controlled by External Electric Field. *J. Phys. Chem. C* **119**, 6912–6917 (2015).
28. Li, X. *et al.* Recent Advances in Noncontact External-Field-Assisted Photocatalysis: From Fundamentals to Applications. *ACS Catal.* **11**, 4739–4769 (2021).
29. Hu, C. *et al.* Photocatalysis Enhanced by External Fields. *Angew. Chem. Int. Ed.* **60**, 16309–16328 (2021).
30. Dai, B. *et al.* Recent Advances in Efficient Photocatalysis via Modulation of Electric and Magnetic Fields and Reactive Phase Control. *Adv. Mater.* **35**, 2210914 (2023).
31. Li, R. *et al.* Research Advances in Magnetic Field-Assisted Photocatalysis. *Adv. Funct. Mater.* **34**, 2316725 (2024).
32. Tan, X. *et al.* Magnetic field enabled spin-state reconfiguration for highly sensitive paper-based photoelectrochemical bioanalysis. *Chem. Eng. J.* **471**, 144626 (2023).
33. Chen, Y. *et al.* Ultrasensitive Paper-Based Photoelectrochemical Biosensor for Acetamiprid Detection Enabled by Spin-State Manipulation and Polarity-Switching. *Anal. Chem.* **96**, acs.analchem.4c01251 (2024).
34. Yang, W. *et al.* Reducing Intrinsic Carrier Recombination in Au/CuTCPP(Fe) Schottky Junction Through Spin Polarization Manipulation for Sensitive Photoelectrochemical Biosensing. *Anal. Chem.* **97**, 3756–3764 (2025).
35. Shan, L. *et al.* Spin-State Reconfigurable Magnetic Perovskite-Based Photoelectrochemical Sensing Platform for Sensitive Detection of Acetamiprid. *Adv. Funct. Mater.* 2418023 (2025).
36. Li, J. *et al.* Enhanced Photocatalytic Performance through Magnetic Field Boosting Carrier Transport. *ACS Nano* **12**, 3351–3359 (2018).
37. Xie, Q. *et al.* Field-free magnetization switching induced by the unconventional spin–orbit torque from $WTe_2$. *APL Mater.* **9**, 051114 (2021).
38. He, J. *et al.* Photocatalytic $H_2O$ Overall Splitting into $H_2$ Bubbles by Single Atomic Sulfur Vacancy CdS with Spin Polarization Electric Field. *ACS Nano* **15**, 18006–18013 (2021).
39. Xie, Q. *et al.* Giant Enhancements of Perpendicular Magnetic Anisotropy and Spin-Orbit Torque by a $MoS_2$ Layer. *Adv. Mater.* **31**, 1900776 (2019).
40. Binasch, G., Grünberg, P., Saurenbach, F. & Zinn, W. Enhanced magnetoresistance in layered

magnetic structures with antiferromagnetic interlayer exchange. *Phys. Rev. B* **39**, 4828–4830 (1989).

41. Chappert, C. The emergence of spin electronics in data storage. *Nat. Mater.* **6**, 813–823 (2007).
42. Lin, C.-C. *et al.* Spin-Polarized Photocatalytic $CO_2$ Reduction of Mn-Doped Perovskite Nanoplates. *J. Am. Chem. Soc.* **144**, 15718–15726 (2022).
43. Gao, W. *et al.* Electron Spin Polarization-Enhanced Photoinduced Charge Separation in Ferromagnetic $ZnFe_2O_4$. *ACS Energy Lett.* **6**, 2129–2137 (2021).
44. Zhou, T. *et al.* Mechanisms and recent advances in external magnetic field assisted photocatalysis: A mini review. *Mater. Today Commun.* **44**, 111969 (2025).
45. Sun, Y., Zhang, Y., Zhang, H., Liu, M. & Liu, Y. Integrating Highly Efficient Recognition and Signal Transition of $g-C_3N_4$ Embellished $Ti_3C_2$ MXene Hybrid Nanosheets for Electrogenerated Chemiluminescence Analysis of Protein Kinase Activity. *Anal. Chem.* **92**, 10668–10676 (2020).
46. Halim, J. *et al.* X-ray photoelectron spectroscopy of select multi-layered transition metal carbides (MXenes). *Appl. Surf. Sci.* **362**, 406–417 (2016).
47. Ye, C. *et al.* $Ti_3C_2$ MXene-based Schottky photocathode for enhanced photoelectrochemical sensing. *J. Alloys Compd.* **859**, 157787 (2021).
48. Chen, F., Wang, J., Chen, L., Lin, H. & Han, D. A Wearable Electrochemical Biosensor Utilizing Functionalized $Ti_3C_2T_x$ MXene for the Real-Time Monitoring of Uric Acid Metabolite. *Anal Chem* **96**, 3914–3924 (2024).
49. Meng, C. *et al.* Angstrom-confined catalytic water purification within $Co-TiO_x$ laminar membrane nanochannels. *Nat. Commun.* **13**, 4010 (2022).
50. Li, M. *et al.* Enhanced Salinity Gradient Energy Conversion by Photodegradable $MXene/TiO_2$ Membrane Utilizing Saline Dye Wastewater. *Adv. Funct. Mater.* **35**, 2414342 (2025).
51. Gao, W. *et al.* Electromagnetic induction derived micro-electric potential in metal-semiconductor core-shell hybrid nanostructure enhancing charge separation for high performance photocatalysis. *Nano Energy* **71**, 104624 (2020).
52. Wang, H., Zhang, B., Zhao, F. & Zeng, B. One-Pot Synthesis of N-Graphene Quantum Dot-Functionalized I-BiOCl Z-Scheme Cathodic Materials for "Signal-Off" Photoelectrochemical Sensing of Chlorpyrifos. *ACS Appl. Mater. Interfaces* **10**, 35281–35288 (2018).
53. Zhang, H., Wang, Z., Zhang, Q., Wang, F. & Liu, Y. $Ti_3C_2$ MXenes nanosheets catalyzed highly efficient electrogenerated chemiluminescence biosensor for the detection of exosomes. *Biosens. Bioelectron.* **124–125**, 184–190 (2019).
54. Zhang, C. *et al.* Mass production of 2D materials by intermediate-assisted grinding exfoliation. *Natl. Sci. Rev.* **7**, 324–332 (2020).
55. Ding, B. *et al.* Giant magneto-birefringence effect and tuneable colouration of 2D crystal suspensions. *Nat. Commun.* **11**, 3725 (2020).

# Supporting information for Magneto-photoelectrochemical 2D heterojunction platform for biosensing detection

Tao Wang[1#], Nan Zhang[1#*], Hongjie Huang[4], Yunhe An[2], Yunyun Dai[5], Yongrui Li[1], Nan Yang[2], Chaojie Yang[2], Xinran Zhou[2], Yucheng Zhu[4], Yingshan Ma[5], Lingling Huang[1*], Yongtian Wang[1], Yang Liu[3*], Zhiyong Yan[2*]

[1]Beijing Engineering Research Center of Mixed Reality and Advanced Display, School of Optics and Photonics, Beijing Institute of Technology, Beijing, 100081, China.
[2]Institute of Analysis and Testing, Beijing Academy of Science and Technology (Beijing Center for Physical and Chemical Analysis), Beijing, 100089, China.
[3]Department of Chemistry, Kay Lab of Bioorganic Phosphorus Chemistry and Chemical Biology of Ministry of Education, Beijing Key Laboratory for Analytical Methods and Instrumentation, Tsinghua University, Beijing, 100084, China.
[4]Department of Sports Medicine Peking University Third Hospital, Institute of Sports Medicine of Peking University, Beijing Key Laboratory of Sports Injuries, Engineering Research Center of Sports Trauma Treatment Technology and Devices Ministry of Education, Beijing, 100191, China.
[5]School of Integrated Circuits and Electronics, Beijing Institute of Technology, Beijing, 100081 China.

[#]These authors contributed equally: Tao Wang, Nan Zhang
[*]Corresponding Authors e-mail: nanzhang@bit.edu.cn; huanglingling@bit.edu.cn; liu-yang@mail.tsinghua.edu.cn; yanzhiyong@bcpca.ac.cn

**This file includes:**

Supplementary Note 1: Formulas for calculating conduction band potentials and valence band potentials

Supplementary Note 2: Computational details

Fig. S1 SEM image of the MXenes film.

Fig. S2 AFM image of the Co-TiO$_2$ Nanosheet.

Fig. S3 EDX mapping of the elemental composition of MXenes.

Fig. S4 EDX mapping of the elemental composition of Co-TiO$_2$.

Fig. S5 XPS spectra of MXenes powder.

Fig. S6 Preparation process of biosensor.

Fig. S7 The photocurrent response diagram corresponding to progressively modified electrodes.

Fig. S8 PL spectra of Co-TiO$_2$ with MF and without MF conditions.

Fig. S9 The intensity of the photocurrent signal changes with different kempatide concentrations.

Fig. S10 The linear relationship between the intensity of the photocurrent changes and the PKA concentration without magnetic field.

Fig. S11 The detection of PKA concentration in the synovial fluid (diluted 500 times) of normal individuals and osteoarthritis patients using an Elisa kit.

Fig. S12 8-cycle response curve of the PEC biosensor.

Table 1 Comparison of various methods for PKA activity assay

**Supplementary Note 1: Formulas for calculating conduction band potentials and valence band potentials.**

The conversion equation normal hydrogen electrode potential (NHE) and Ag/AgCl electrode potential[1]:

$$E_{CB}(\text{V vs Ag/AgCl}) = E_{fb}(\text{V vs Ag/AgCl}) - 0.2 \text{ V}$$

$$E_{CB}(\text{V vs NHE}) = E_{CB}(\text{V vs Ag/AgCl}) + 0.24 \text{ V}$$

$$E_{VB}(\text{V vs NHE}) = E_g(\text{eV}) + E_{CB}(\text{V vs Ag/AgCl})$$

**Supplementary Note 2: Computational details.**

All the calculations were carried out using the GGA-PBE functional implemented in the Vienna ab initio Simulation Package (VASP)[2]. We used the generalized gradient approximation (GGA) with the Pedrew-Burke-Ernzerhof (PBE) function[3] and Hubbard U corrections (3.32 eV for treating Co 3d orbitals) were introduced to consider the self-interaction error of transition metals. The cut-off energy for plane wave is set to 500 eV. The energy criterion is set to 10-5 eV in iterative solution of the Kohn-Sham equation. The Brillouin zone integration is performed using a 3x3x1 k-mesh. All the structures are relaxed until the residual forces on the atoms have declined to less than 0.01 eV/Å.

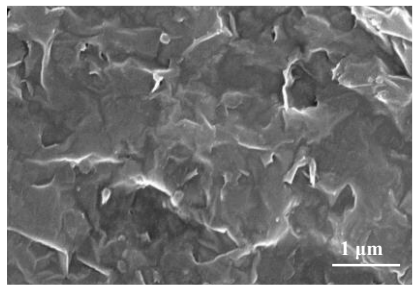

**Fig. S1 SEM image of the MXenes film.**

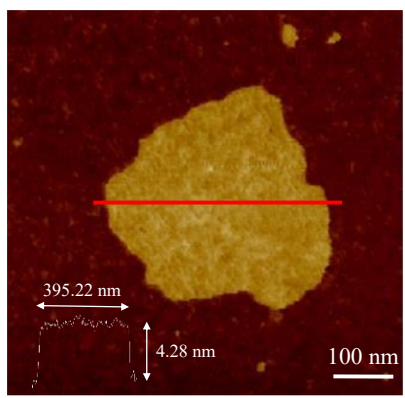

**Fig. S2 AFM image of the Co-TiO$_2$ Nanosheet.** The Co-TiO$_2$ Nanosheet showed a thickness of 4.28 nm and a lateral size of 395.22 nm.

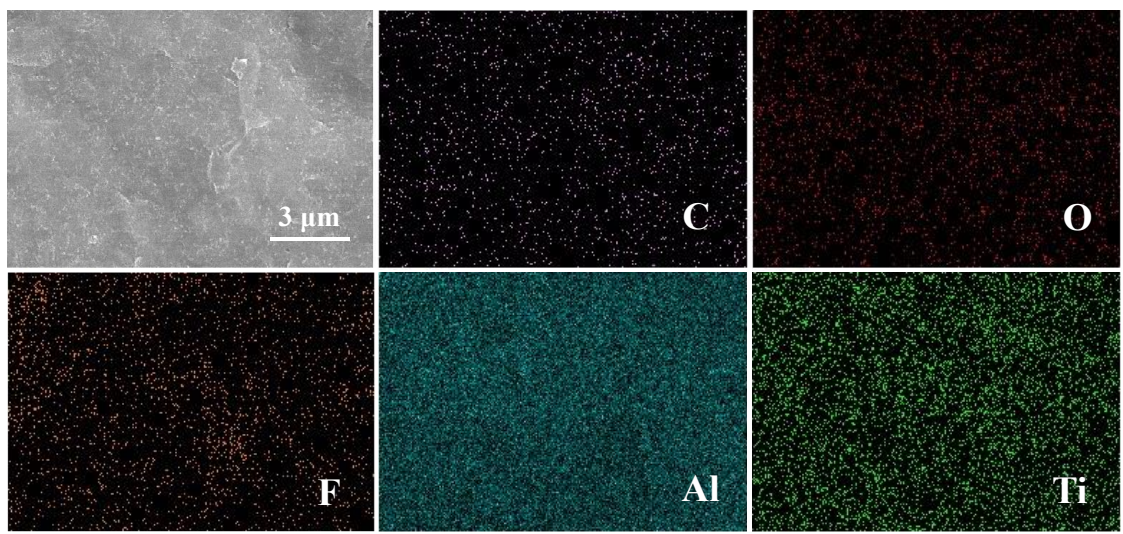

**Fig. S3 EDX mapping of the elemental composition of MXenes.** MXenes contained C, O, F, Al and Ti elements.

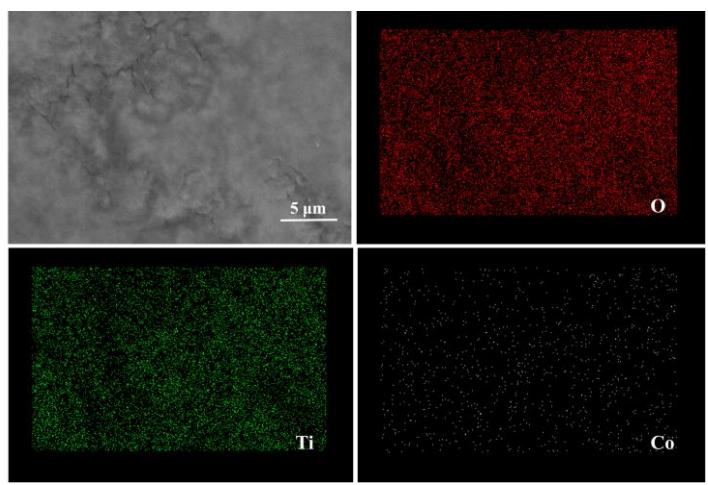

**Fig. S4 EDX mapping of the elemental composition of Co-TiO$_2$.** Co-TiO$_2$ contained O, Ti and Co elements.

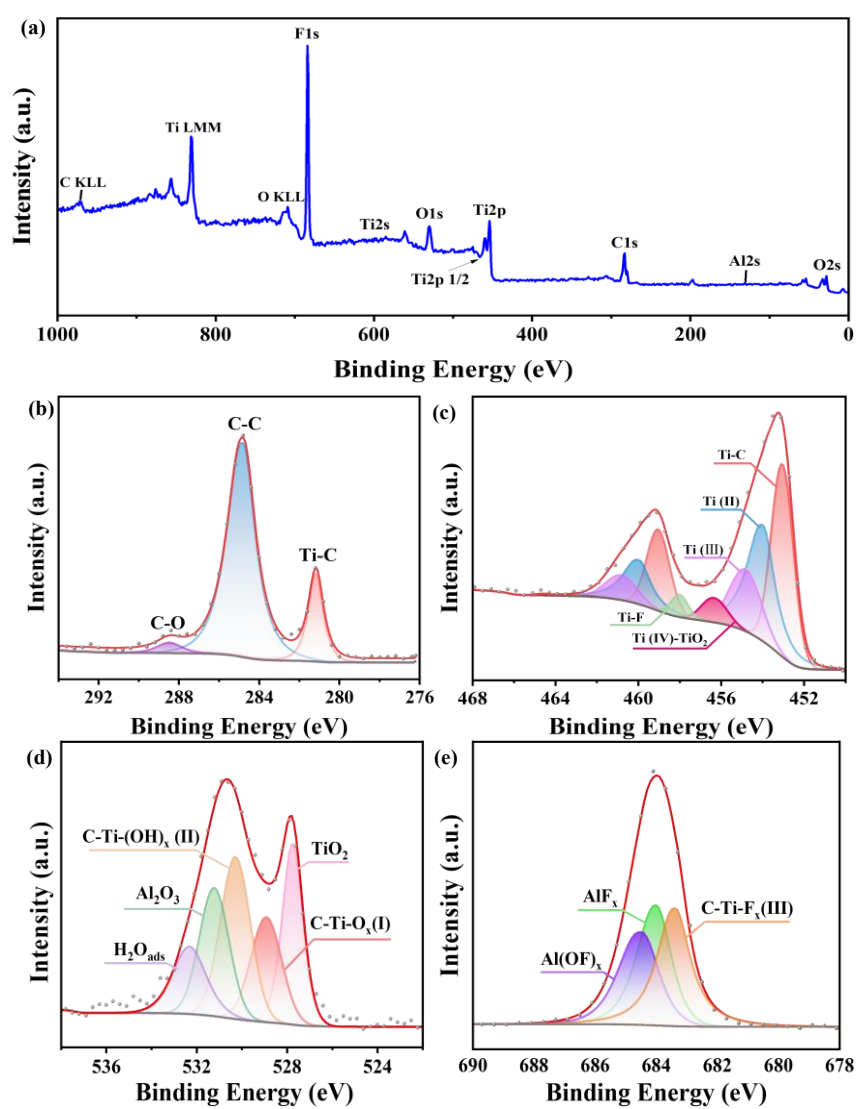

**Fig. S5 XPS spectra of MXenes powder.** (a) survey, (b–e) high resolution spectra of C1s, Ti 2p, O 1s, and F1s regions, respectively.

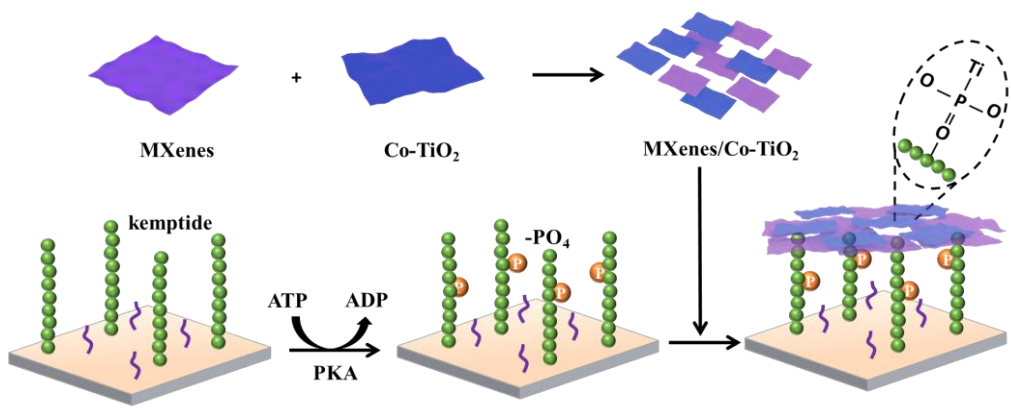

**Fig. S6 Preparation process of biosensor.** In brief, Ti$^{4+}$ sites within the MXenes/Co-TiO$_2$ composite selectively recognize phosphate groups on the kinase phosphorylation substrate (kemptide) via coordination interactions. Detailed descriptions are provided in the main text.

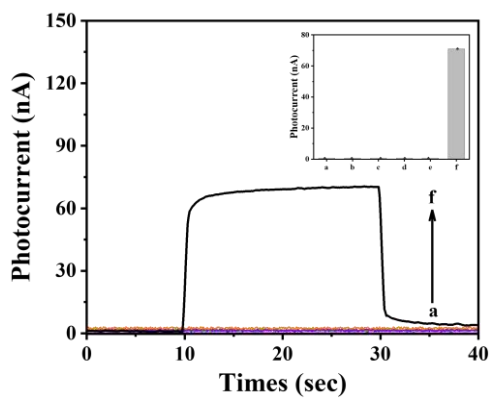

**Fig. S7 The photocurrent response diagram corresponding to progressively modified electrodes.** (a) bare ITO, (b) chitosan-modified ITO, (c) glutaraldehyde/chitosan/ITO, (d) kemptide/glutaraldehyde/chitosan/ITO, (e) phosphorylated kemptide/glutaraldehyde/chitosan/ITO, and (f) MXenes/Co-TiO$_2$ probe-functionalized electrode.

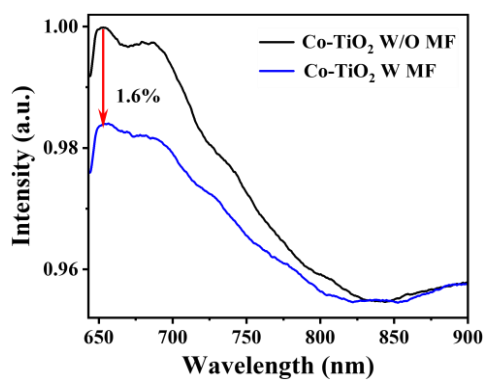

**Fig. S8 PL spectra of Co-TiO$_2$ with MF and without MF conditions.** Under 532 nm laser excitation, the PL intensity of Co-TiO$_2$ at 650 nm decreased by 1.6% under MF.

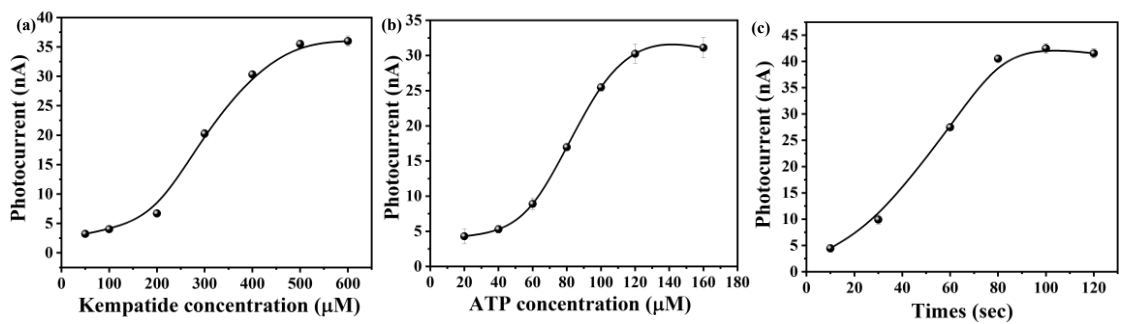

**Fig. S9 The intensity of the photocurrent signal changes with different kempatide concentrations.** (a) kempatide concentration, (b) ATP concentrationsand (c) phosphorylation timesc. The light power density was 190 mW cm$^{-2}$, electrode area was 0.5 cm$^2$.

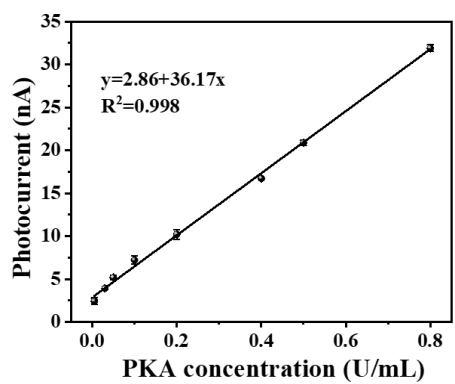

**Fig. S10 The linear relationship between the intensity of the photocurrent changes and the PKA concentration without magnetic field.** The PKA concentration within the range of 0.005 to 1 U/mL had a linear fitting function of y = 2.86 + 36.17x (where y represents photocurrent intensity and x represents PKA concentration), with a correlation coefficient of $R^2$ = 0.998.

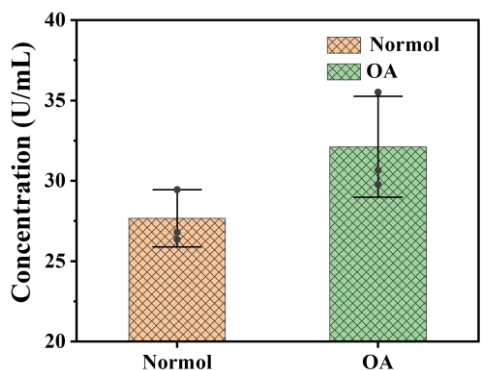

**Fig. S11 The detection of PKA concentration in the synovial fluid (diluted 500 times) of normal individuals and osteoarthritis patients using an Elisa kit.** The results indicated that the biological activity of PKA in the synovial fluid of osteoarthritis patients is significantly increased

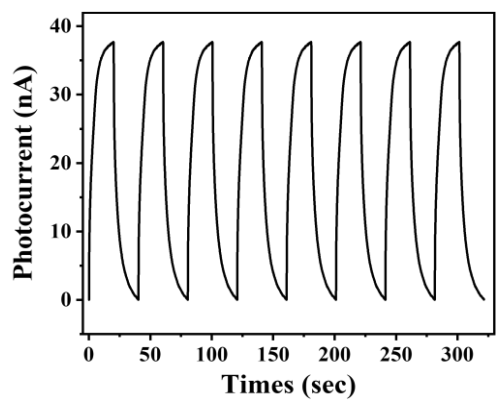

**Fig. S12 8-cycle response curve of the PEC biosensor.** The PKA concentration was 0.5 U/mL. The results showed a low relative standard deviation (RSD) of 0.33%.

Table 1 Comparison of various methods for PKA activity assay

| Method | Linear range (U/mL) | Detection limit (U/mL) | Reference |
|---|---|---|---|
| Colorimetry | 0-0.5 | 0.0375 | 4 |
| Fluorescence | 0-1000 | 0.5 | 5 |
| Fluorescence | 0-800 | 0.5 | 6 |
| Electrochemiluminescence | 0.07-32 | 0.07 | 7 |
| Quartz crystal | 0.64-22.33 | 0.061 | 8 |
| Electrochemistry | 0.05-100 | 0.014 | 9 |
| Electrochemistry | 0.005-50 | 0.002 | 10 |
| PEC method | 0.005-0.0625 | 0.0049 | 11 |
| PEC method | 0.05-100 | 0.048 | 12 |
| PEC method | 0.05-50 | 0.017 | 13 |
| PEC method | 0.001-100 | 0.00035 | 14 |
| PEC method | 0.008-80 | 0.005 | 15 |
| PEC method | 0.005-50 | 0.0049 | 16 |
| PEC method | 0.015-60 | 0.009 | 17 |
| PEC method | 0.004-90 | 0.0026 | 18 |
| This work | 0.005-80 | 0.00027 | without MF |
| This work | 0.005-80 | 0.00016 | With MF |

**References**

1. Li, M. *et al.* Enhanced Salinity Gradient Energy Conversion by Photodegradable MXene/TiO$_2$ Membrane Utilizing Saline Dye Wastewater. *Adv. Funct. Mater.* **35**, 2414342 (2025).
2. Perdew, J. P., Burke, K. & Ernzerhof, M. Generalized Gradient Approximation Made Simple. *Phys. Rev. Lett.* **77**, 3865–3868 (1996).
3. Kresse, G. & Joubert, D. From ultrasoft pseudopotentials to the projector augmented wave method. *Phys. Rev. B* **59**, 1758–1775 (1999).
4. Shen, C. *et al.* Colorimetric and electrochemical determination of the activity of protein kinase based on retarded particle growth due to binding of phosphorylated peptides to DNA – capped silver nanoclusters. *Microchim. Acta* **183**, 2933–2939 (2016).
5. Lu, G. *et al.* Fluorescent detection of protein kinase based on positively charged gold nanoparticles. *Talanta* **128**, 360–365 (2014).
6. Tan, P. *et al.* Fluorescent detection of protein kinase based on zirconium ions immobilized magnetic nanoparticles. *Anal. Chim. Acta* **780**, 89–94 (2013).
7. Xu, S., Liu, Y., Wang, T. & Li, J. Highly Sensitive Electrogenerated Chemiluminescence Biosensor in Profiling Protein Kinase Activity and Inhibition Using Gold Nanoparticle as Signal Transduction Probes. *Anal. Chem.* **82**, 9566–9572 (2010).
8. Xu, X. *et al.* Aptameric Peptide for One-Step Detection of Protein Kinase. *Anal. Chem.* **84**, 4746–4753 (2012).
9. Zhou, Y. *et al.* Electrochemical biosensor for protein kinase A activity assay based on gold nanoparticles-carbon nanospheres, phos-tag-biotin and β-galactosidase. *Biosens. Bioelectron.* **86**, 508–515 (2016).
10. Shen, C. *et al.* A single electrochemical biosensor for detecting the activity and inhibition of both protein kinase and alkaline phosphatase based on phosphate ions induced deposition of redox precipitates. *Biosens. Bioelectron.* **85**, 220–225 (2016).
11. Wang, Z. *et al.* Highly sensitive photoelectrochemical biosensor for kinase activity detection and inhibition based on the surface defect recognition and multiple signal amplification of metal-organic frameworks. *Biosens. Bioelectron.* **97**, 107–114 (2017).
12. Li, X., Zhu, L., Zhou, Y., Yin, H. & Ai, S. Enhanced Photoelectrochemical Method for Sensitive Detection of Protein Kinase A Activity Using TiO$_2$/g-C$_3$N$_4$, PAMAM Dendrimer, and Alkaline Phosphatase. *Anal. Chem.* **89**, 2369–2376 (2017).
13. Sui, C. *et al.* Photoelectrochemical determination of the activity of protein kinase A by using g-C$_3$N$_4$ and CdS quantum dots. *Microchim. Acta* **185**, 541 (2018).
14. Xiao, K. *et al.* A label-free photoelectrochemical biosensor with near-zero-background noise for protein kinase A activity assay based on porous ZrO$_2$/CdS octahedra. *Sens. Actuators B Chem.* **328**, 129096 (2021).
15. Yan, Z., Wang, Z., Miao, Z. & Liu, Y. Dye-Sensitized and Localized Surface Plasmon Resonance Enhanced Visible-Light Photoelectrochemical Biosensors for Highly Sensitive Analysis of Protein Kinase Activity. *Anal. Chem.* **88**, 922–929 (2016).
16. Wang, Z. Highly sensitive photoelectrochemical biosensor for kinase activity detection and inhibition based on the surface defect recognition and multiple signal amplification of metal-organic frameworks. *Biosens. Bioelectron.* **97**, 107–114 (2017).
17. Yan, Z. Sensitive electrogenerated chemiluminescence biosensors for protein kinase

activity analysis based on bimetallic catalysis signal amplification and recognition of Au and Pt loaded metal-organic frameworks nanocomposites. *Biosens. Bioelectron.* **109**, 132–138 (2018).
18. Yan, Z. *et al.* Sensitive photoelectrochemical biosensors based on AuNPs/MXenes electrode coupled with light-harvesting UiO-66-NH$_2$ probes for protein kinase detection. *Biosens. Bioelectron. X* **11**, 100204 (2022).